\documentclass[aps,prb,twocolumn,superscriptaddress,showpacs]{revtex4}

\usepackage{latexsym} 
\usepackage{graphicx}
\usepackage[applemac]{inputenc} 
\usepackage{amsmath}
\usepackage{bm}

\begin{document}

\title{Subtle competition between ferromagnetic and antiferromagnetic  order in a Mn(II) - free radical ferrimagnetic chain}

\author{E. Lhotel}
\affiliation{Centre de Recherche sur les Tr\`es Basses
Temp\'eratures, CNRS, BP~166, 38042 Grenoble, France}
\author{V. Simonet}
\email[Corresponding author: ]{virginie.simonet@grenoble.cnrs.fr}
\affiliation{Laboratoire Louis N\'eel, CNRS, BP~166, 38042
Grenoble, France}
\author{E. Ressouche}
\affiliation{CEA-Grenoble, DRFMC / SPSMS / MDN, 17 rue des
Martyrs, F-38054 Grenoble cedex, France}
\author{B. Canals}
\affiliation{Laboratoire Louis N\'eel, CNRS, BP~166, 38042
Grenoble, France}
\author{D. B. Amabilino}
\author{C. Sporer}
\affiliation{Institut de Ci\`encia de Materials de
Barcelona (CSIC), Campus Universitari, 08193 Bellaterra, Spain}
\author{D. Luneau}
\affiliation{Universit\'e Claude Bernard-Lyon 1, Laboratoire des Multimat\'eraiux et Interfaces (UMR 5615), B\^at Raulin - Campus de la Doua, 69622 Villeurbanne Cedex, France}
\author{J. Veciana}
\affiliation{Institut de Ci\`encia de Materials de
Barcelona (CSIC), Campus Universitari, 08193 Bellaterra, Spain}
\author{C. Paulsen}
\affiliation{Centre de Recherche sur les Tr\`es Basses
Temp\'eratures, CNRS, BP~166, 38042 Grenoble, France}

\begin{abstract}
The macroscopic magnetic characterization of the Mn(II) - nitronyl
nitroxide free radical chain (Mn(hfac)$_2$({\it R})-3MLNN)
evidenced its transition from a 1-dimensional behavior of
ferrimagnetic chains to a 3-dimensional ferromagnetic long range
order below 3 K. Neutron diffraction experiments, performed on a
single crystal around the transition temperature, led to a
different conclusion : the magnetic Bragg reflections detected
below 3 K correspond to a canted antiferromagnet where the
magnetic moments are mainly oriented along the chain axis.
Surprisingly in the context of other compounds in this family of
magnets, the interchain coupling is antiferromagnetic. This state
is shown to be very fragile since a ferromagnetic interchain
arrangement is recovered in a weak magnetic field. This peculiar
behavior might be explained by the competition between dipolar
interaction, shown to be responsible for the antiferromagnetic
long range order below 3 K, and exchange interaction, the balance
between these interactions being driven by the strong intrachain
spin correlations. More generally, this study underlines the need,
in this kind of molecular compounds, to go beyond macroscopic
magnetization measurements.
\end{abstract}


\maketitle

\section{Introduction}

The organic chemistry synthesis approach, applied in the field of
molecular magnetism, has allowed to obtain original compounds with
properties characteristic of both classical magnets and organic
compounds. Some of the early successes reported in this field were
the finding of a purely organic ferromagnet \cite{Kinoshita91},
the experimental proof of the Haldane conjecture in a molecular
chain compound \cite{Renard88}, or the study of single molecule
magnets providing model systems for the investigation of quantum
effects such as quantum tunneling of magnetization or topological
interferences \cite{Thomas96, Wernsdorfer99}.

Another important achievement was made in the field of molecular
magnetic one-dimensional (1D) compounds with the design of
ferrimagnetic chain compounds. They consist of magnetically
quasi-isolated chains where two kinds of different magnetic
centers alternate regularly. The first example was identified in a
bimetallic ferrimagnetic chain compound \cite{Verdaguer84} built
of Mn$^{2+}$ ions with spin 5/2 and Cu$^{2+}$ ones with spin 1/2.
Several other types of ferrimagnetic chains were then discovered
with various metallic centers, different spatial intra and
interchain architectures, alternating interaction pathways within
the chains, leading to different magnetic properties especially at
low temperature where interchain interactions begin to play a role
\cite{Kahn}. Not only bimetallic materials but also
metallo-organic chains, in which the spin carriers are a metallic
ion and an organic free radical, generally nitroxide, were synthesized and proved to enrich
further the variety of magnetic behaviors
\cite{Kahn,CaneschiCu,CaneschiNi,CaneschiMn,CaneschiCo, Luneau1, Luneau2}. Very
recently, this kind of materials was reinvestigated with new
questions concerning in particular the nature of the slow spin
dynamics in Ising like chain compounds
\cite{Caneschi02,Coulon04,Wernsdorfer05}. The slow relaxation of
magnetization, in first order agreement with the theory of Glauber
concerning Ising 1D chain \cite{Glauber63}, is a consequence of
strong uniaxial anisotropy and also of intrachain magnetic
correlations.

The success of the metallo-organic route chosen by several groups
hold on the use of nitronyl
nitroxide free radicals (NITR) where R stands for an alkyl or
aromatic group. This organic entity has one unpaired electron
carrying a spin 1/2, delocalized on the 5 atoms of its O-N-C-N-O
fragment as shown by polarized neutron studies on a compound of
very weakly interacting nitronyl nitroxide molecules
\cite{Zheludev94}. This original electronic configuration allows
the radical to be coupled simultaneously to two metallic groups.
These groups can be built of one magnetic 3d metallic ion M like
Cu, Ni, Mn or Co, surrounded by magnetic inactive bulky organic
moieties like hexafluoroacetylacetonates (hfac)
\cite{CaneschiCu,CaneschiNi,CaneschiMn,CaneschiCo,Luneau2, Minguet02}. The
use of this NITR free radical leads preferentially to
antiferromagnetic intrachain coupling stronger than in bimetallic
ferrimagnetic chain compounds, except in the case of some Cu
complexes where it is ferromagnetic. The possible cis or trans
NITR-M(hfac)$_2$ coordination scheme, determined by the R group,
can produce different architectural arrangements leading to
linear, zig-zag or helical chains with different magnetic
properties. These were studied by Caneschi {\it et al.}
\cite{CaneschiMn} within the rich family of the Mn-NITR compounds.
Several compounds with different R group, studied by magnetometry
and EPR, were shown to present a ferrimagnetic 1D behaviour. At
low temperature, between 5 and 9 K, they undergo a transition
towards a three-dimensional (3D) long-range order (LRO) consisting
of a ferromagnetic ordering of the Mn-NITR entities.

In these materials, chains are well magnetically isolated from
each other thanks to their large separation by bulky magnetically
inactive organic moieties. This results in very weak interchain
exchange interactions, leading to a ratio of intra to interchain
interaction that reaches several orders of magnitude and a strong
1D behaviour. Then it is necessary to take into account other
usually neglected interactions such as the dipolar one. In pure 1D
compound, LRO is not expected at finite temperature, but very weak
interchain dipolar interactions, reinforced by strong intrachain
correlations, can induce such a 3D LRO
\cite{Walker,Wynn,Ostrovsky01}. The role of dipolar interaction is
also invoked to explain the 3D magnetic ordering of high spin
molecular cluster compounds \cite{Morello,Evangelisti}.

\begin{figure}
\includegraphics[width=7cm]{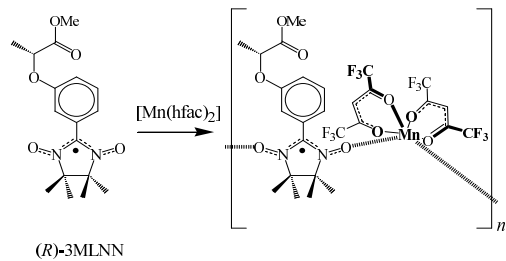}
\caption{Schematic representation of the (Mn(hfac)$_2$({\it
R})-3MLNN complex.} \label{complex}
\end{figure}

The present studied chain compound \cite{Minguet02},
Mn(hfac)$_2$({\it R})-3MLNN with the NITR free radical ({\it
R})-3MLNN=({\it
R})-Methyl[3-(4,4,5,5-tetramethyl-4,5-dihydro-1{\it
H}-imidazolyl-1-oxy-3-oxide) phenoxy]-2-propionate, shown in Fig.~\ref{complex}, is formed by alternating Mn$^{2+}$ ions carrying a
spin 5/2 and chiral nitronyl nitroxide free radicals. It
crystallizes in the non-centrosymmetric orthorhombic
P2$_1$2$_1$2$_1$ space group. There are 4 formulas
C$_{27}$H$_{25}$F$_{12}$Mn$_2$O$_9$ per unit cell which sum up to
304 atoms. The Mn$^{2+}$ ions are in the center of a distorted
oxygen octahedron, linked to two hfac moieties and two NITR
radicals in cis coordination which gives the compound its zig-zag
structure. The resulting chains propagate along the $b$ axis with
Mn-Mn distances of 7.57 \AA~ within the chains and of at least
11.3 \AA~ between adjacent chains. Each chain is surrounded by 6
others, some of those related by a screw axis.

The possibility to work with a Mn(hfac)$_2$({\it R})-3MLNN
single-crystal, sufficiently stable for neutron diffraction
measurements, allowed us for the first time in this kind of
metallo-organic ferrimagnetic chain, to determine unambiguously
the nature of the 3D LRO stabilized at low temperature and to
evidence strong magnetic competition. A first magnetic
investigation of this compound by macroscopic measurements was
reported in Ref. \onlinecite{Minguet02}, and is recalled in section II.
The determination by neutron diffraction of the magnetic LRO below
the transition is presented in section III.1. Since it can not be
explained with the same kind of magnetic interaction than the
correlations probed by magnetic measurements, a neutron
diffraction experiment under a magnetic field was further
performed (cf. section III.2) that showed the low robustness of
the LRO state versus a very small field. Those results are
discussed in terms of competing interactions in section IV.

\section{Magnetization measurements}

The macroscopic magnetic properties of Mn(hfac)$_2$({\it R})-3MLNN
single-crystals were investigated by dc magnetization and ac
susceptibility measurements using three superconducting quantum
interference device (SQUID) magnetometers. One is commercial (0-80
kOe, 2-200 K). The two others, developed at the CRTBT/CNRS, are
equipped with a miniature dilution refrigerator: one is devoted to
low field measurements (0-2 kOe), and the second to high field
(0-80 kOe).

Measurements in the three directions have shown that there is a
weak uniaxial anisotropy along the chain direction ($b$ axis). The
anisotropy constant $K \approx 2.5 \times 10^3$~J.m$^{-3}$ was
estimated from saturation magnetization and perpendicular
susceptibility measurements at low temperatures. Hereafter, we
will focus on measurements performed along the easy axis.

The high temperature characterization of the sample was made by
magnetization measurements  in a field of  H$_{dc}$=10 Oe from
T=20-90 K and  by AC susceptibility measurements  in a field
H$_{ac}$=1.4~Oe from T=5-20~K. Throughout these  temperature
ranges, the fields are small enough to ensure that the
magnetization is linear in field, so that the magnetization M/H
equals the linear susceptibility $\chi$.

\begin{figure}
\includegraphics[width=7cm]{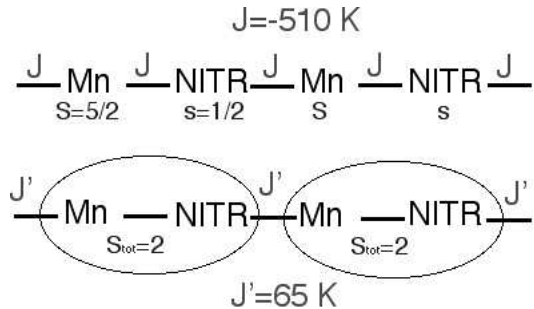}
\caption{Schematic representation of the intrachain interactions J
and J'.} \label{schemaJ}
\end{figure}

A 1D ferrimagnetic behavior expected from the non-compensation of
the Mn$^{2+}$ isotropic spins S=5/2 and the NITR free radical
spins s=1/2 is observed above 20~K \cite{Minguet02}. The high
temperature magnetic susceptibility is indeed compatible with a
strong antiferromagnetic coupling $J/k_B$=-510 K between the Mn
and the NITR spins inside the chains \cite{Seiden83,footnote1}
which is not broken even at room temperature \cite{Minguet02}.
Therefore, at lower temperature (below 100 K), the chains can be
considered as $S_{tot}$=S-s=2 effective units, ferromagnetically
coupled by an exchange $J'$ as represented in Fig.~\ref{schemaJ}.
Between 20 and 90~K ($J'S_{tot}^2/k_BT \gg 1$), the susceptibility
$\chi$(T) can then be fitted in the framework of 1D ferromagnetic
chains \cite{Suzuki94} by Eq.~\ref{Eqchi1D}. The fit yields
$J'/k_B$=65 K.

\begin{equation}
20\ K < T < 90\ K, \quad  \chi=\chi_{1D}= \alpha
\frac{2}{3J'}\left(\frac{J'S_{tot}^2}{k_B T}\right)^2
\label{Eqchi1D}
\end{equation}
with $\alpha=ng^2 \mu_B^2$, where $n$ is the number of magnetic
entities and setting $g$=2 (value obtained by EPR measurements
\cite{Vidal04}) .

\begin{figure}
\includegraphics[width=7.5cm]{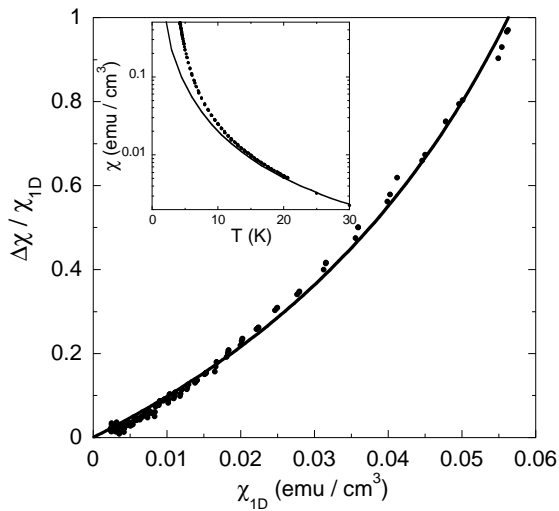}
\caption{$\Delta \chi / \chi_{1D}$ \textit{vs.} $\chi_{1D}$ (see
text) for $5\ K < T < 20\ K$. The line is a fit to Eq.
\ref{Eqdeltachi} with $J_{inter}/k_B=$4.3~mK. The inset shows the
measured $\chi$ \textit{vs.} T (dots) and its departure from the
calculated one (line) for a 1D behavior (Eq. \ref{Eqchi1D}) below
20~K.} \label{figchi}
\end{figure}

Below 20 K, we observe deviations from the 1D susceptibility
$\chi_{1D}$(T) extrapolated using the J' value deduced above 20~K
(see the inset of Fig. \ref{figchi}). This feature announces a
crossover from high temperature 1D to 3D short-range correlations.
The positive difference of $\Delta \chi=\chi-\chi_{1D}$ indicates
the presence of ferromagnetic interactions between the chains. We
can estimate these interactions from a mean-field approach with
the equation \cite{Scalapino75}:

\begin{equation}
5\ K< T < 20\ K, \qquad \frac{\Delta \chi }{\chi_{1D}}= \frac{z
J_{inter}\chi_{1D}/\alpha }{1-z J_{inter}\chi_{1D}/\alpha } \\
\label{Eqdeltachi}
\end{equation}

where $z=6$ is the number of first neighbor chains and $J_{inter}$
is the total interchain exchange interaction. Note that
demagnetization effects are negligible in this temperature range.
The fit between 5 and 20 K, shown in Fig. \ref{figchi}, leads to
$J_{inter}/k_B \approx $~4 mK, which confirms the strong 1D
character of the system: $J_{intra}/J_{inter} > 10^4$.

When the temperature is further decreased, the Mn-radical chain
has been shown to undergo  a magnetic transition at $T_c$=3 K
\cite{Minguet02}. Our analysis of the susceptibility above $T_c$
reveals a small ferromagnetic interchain interaction and in
addition, at $T_c$, the susceptibility exhibits a large peak with
a magnitude  equal to the inverse of the demagnetizing factor N,
within the error bars in the determination of N. As a result of
these observations,   the transition was at first attributed to
ferromagnetic LRO \cite{Minguet02}.

\begin{figure}
\includegraphics[width=7cm]{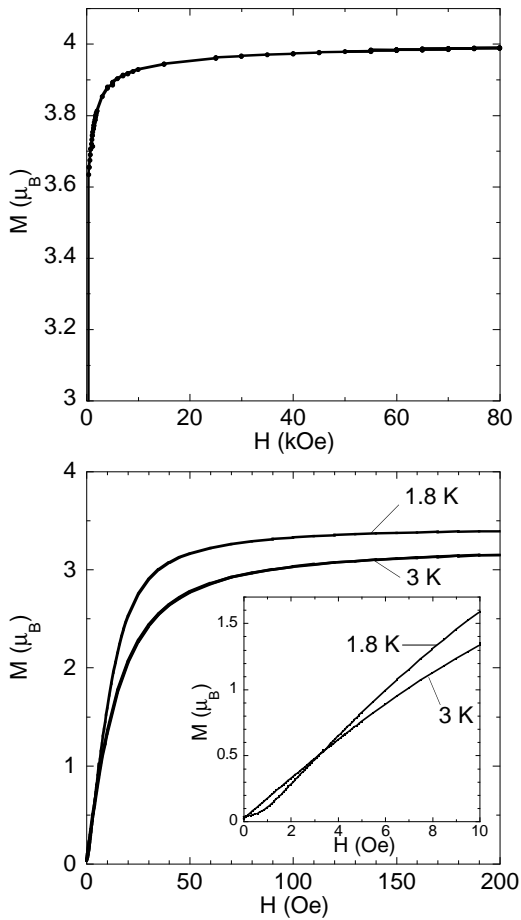}
\caption{$M$ \textit{vs.} $H$ at 200 mK up to 80 kOe (top) ; at
$1.8\ K$ and $3\ K$ up to 200 Oe (bottom) with a zoom in the very
small magnetic field range in the inset.} \label{figMH}
\end{figure}

Indeed, below $T_c$ and in moderated to strong magnetic fields,
the field dependence of the magnetization seems to be well
accounted for by a classical ferromagnetic order. In the high
field regime, magnetization curves saturate rapidly as seen in top
Fig.~\ref{figMH}, although they do not reach the  expected total
saturation value until  80 kOe. In fact, the curves show a non
zero slope which is characteristic of the presence of a small
canting, estimated to be of the order of 10$^{\circ}$ with respect
to the $b$ axis. At low fields, but larger than 5 Oe, the
magnetization curves seem to vary linearly with the field, with a
slope dM/dH approximately equal to the inverse of the
demagnetizing factor (see bottom Fig.~\ref{figMH}). In addition,
below  2.5~K the ac susceptibility is frequency dependent and
presents thermomagnetic irreversibilities, and below 1~K magnetic
hysteresis loops appear as previously reported \cite{Minguet02}.
These properties are reminiscent of ferromagnetic behavior in the
presence of pinning and depinning of domain-walls, and therefore
in agreement with the hypothesis of a canted ferromagnetic order
stabilized in this system.

However, an intriguing behavior is observed for very low fields,
H$<$5~Oe, where the M(H) curves exhibit a negative curvature (see
the curve at 1.8 K in the inset of the bottom Fig.~\ref{figMH}).
Furthermore, the spontaneous magnetization could not be extracted
from the magnetization curves, or from Arrott plots \cite{Stanley}
of the data. The inconsistency in the results, apparent
ferromagnetic behavior without a spontaneous magnetic moment,
suggests that a nonconventional LRO is taking place. This was the
motivation for neutron diffraction measurements in order to
determine the actual magnetic structure of the Mn(hfac)$_2$({\it
R})-3MLNN compound below T$_c$.

\section{Neutron diffraction experiments}

The experiments were performed on the two CRG thermal-neutron
single-crystal diffractometers, D15 and D23, at the
Institut-Laue-Langevin high flux reactor (ILL, Grenoble, France).
Both diffractometers were operated in normal-beam mode, using an
incident wavelength of 1.1694 and 1.2582 \AA~ for D15 and D23
respectively. The D15 experiment was aimed at determining the
nuclear and magnetic structures in zero magnetic field in the
1.5-10 K temperature range around T$_c$. The D23 experiment
yielded information about the evolution of the magnetic structure
under magnetic field and was performed using a vertical cryomagnet
(6 T superconducting coil).

The sample was a very small crystal of Mn(hfac)$_2$({\it R})-3MLNN
of $\approx$ 4.4 mg and 1 mm$^{3}$ which was held in a thin Al box
with apiezon grease and positioned with the $b$ axis vertical on the
spectrometer. This set-up was chosen instead of direct gluing due
to the brittleness of the sample, also very sensitive to
temperature cycling.

\subsection{Nuclear structure}

The first step of the neutron diffraction experiment consisted in
refining the low temperature nuclear structure, previously
determined from X-ray diffraction at room temperature with
calculated H positions. The X-ray structure obtained with a Rf
factor of 19\% suggested the presence of some disorder
\cite{Minguet02}. For the neutron refinement purpose, 352 Bragg
reflections were collected on D15 at 10 K. Unfortunately it was
impossible to collect more reflections because of the fast
deterioration of the sample upon temperature cycling. The
resulting lattice parameters are a=11.84, b=13.79, c=20.03~\AA.
The integrated intensities corrected from the Lorentz factor were
calculated from the $\omega$-scans with the COLL5 program
\cite{Lehmann70}. The averaging of the equivalent reflections and
absorption correction were done using the ARRNGE and AVEXAR
programs from the CCSL library \cite{Brown} with an estimated
total absorption coefficient $\mu$ of 0.12 mm$^{-1}$, essentially
due to the large incoherent scattering of the hydrogen atoms. The
weakness of the measured signal, the presence of important
structural disorder, and the small number of collected Bragg
reflections imposed a limit to the number of free parameters in the
refinement procedure.

\begin{figure*}
\includegraphics[scale=1.2]{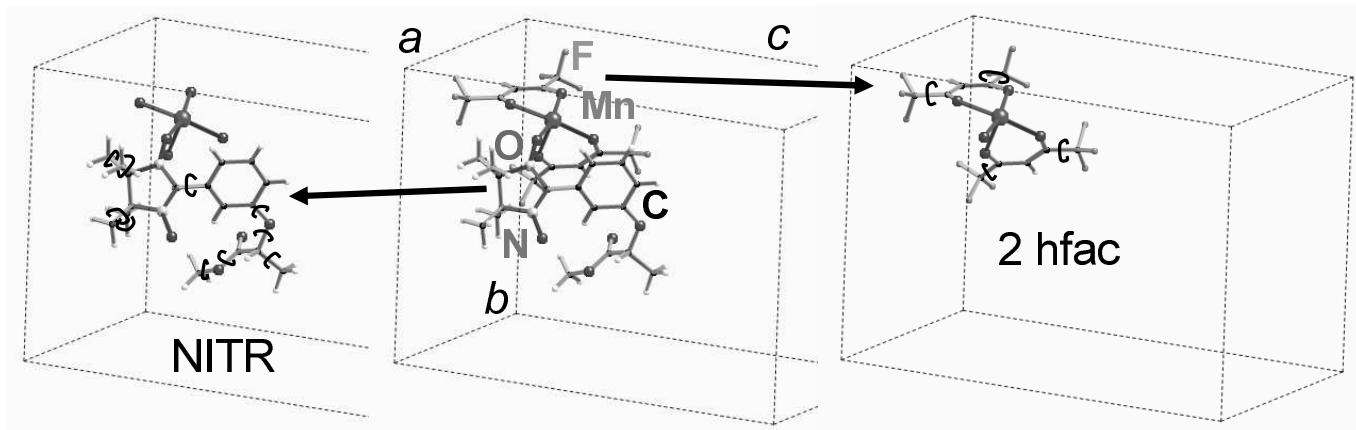}
\caption{Representation of the 13 rigid molecular blocks within
the main NITR and hfac blocks allowed to rotate (black signs
around the rotation axis bond) in the nuclear structure refinement
(see text).} \label{blocks}
\end{figure*}

\begin{figure}
\includegraphics[scale=0.4, bb=0 400 572 814]{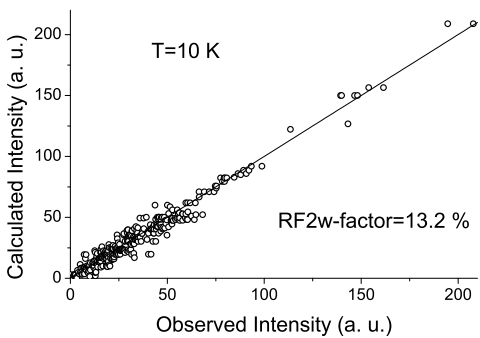}
\caption{Graphical representation of the Mn(hfac)$_2$({\it
R})-3MLNN nuclear structure refinement at 10 K : calculated versus
measured integrated intensities.} \label{Nucl}
\end{figure}

This was achieved using a method based on rigid molecular blocks.
A new implementation in the MXD program \cite{Wolfers90} was used
to handle this by adjusting, for each block of atoms, the 3
coordinates of its origin, and the 3 eularian angles
parameterizing its three dimensional rotation around this point. 3
main blocks were chosen on chemical grounds : the block associated
to the NITR radical had free origin coordinates and free rotation
angles whereas the two hfac blocks had a fixed origin with respect
to the Mn, only the rotation angles being free in the refinement.
13 minor groups (shown in Fig. \ref{blocks}) were further selected
within these main blocks, which were allowed to rotate around the
parent C-C or C-O bond axis: 4 CF$_3$ groups within the hfac
blocks, 5 methyl groups, 1 phenyl group and the 4 segments of the
carbonate chain linked to the phenyl within the NITR block.
Including one isotropic thermal factor for each kind of chemical
species, this led us to fit the measured intensities with 36
refinable variables for 358 observations. The resulting structure,
obtained with a weighted Least Square R factor of 13.2 \% (see
Fig. \ref{Nucl}), presents slight modifications with respect to
the published structure \cite{Minguet02}, consisting essentially
of rigid block rotation angle deviation of 5.3$^{\circ}$ at most.
This treatment provided us with a determination of the positions
of the magnetic atoms and of the scaling factor necessary for a
quantitative analysis of the magnetic structure.

\subsection{Magnetic structure in zero field}

A weak magnetic signal was observed at 1.5 K on top of some
nuclear peaks from the difference with the 10~K measurements and
on the ($h$,0,0) and (0,0,$l$) reflections with odd $h$ and $l$
indices. Those reflections, absent in the nuclear pattern, are
forbidden in this space group due to point group symmetries
relating the 4 asymmetric units. These observations for such a
non-Bravais lattice imply a ${\bm k}$=0 propagation vector with an
antiferromagnetic arrangement of the 4 magnetic entities per unit
cell.

\begin{figure}
\includegraphics[scale=0.4, bb=0 380 572 814]{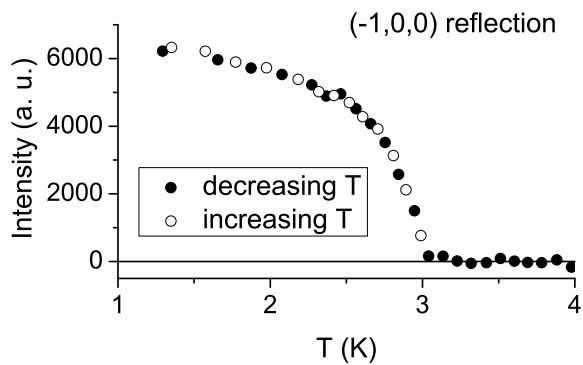}
\caption{Temperature variation of the magnetic signal associated
to the (-1,0,0) reflection : neutron counts at the peak maximum
corrected from the 2.5$^{\circ}$ $\omega$-shifted background. The
error bars are smaller than the symbol size.} \label{MvsT}
\end{figure}

In order to check the intrinsic origin of the observed magnetic
Bragg peaks, the intensity at the peak maximum position for the
purely magnetic (-1,0,0) reflection, characteristic of
antiferromagnetic LRO, was followed with increasing and decreasing
temperature, as shown in Fig. \ref{MvsT}. The signal vanishes at 3
K in agreement with the transition temperature deduced from the
susceptibility measurements. Note the presence of a small kink in
the magnetization versus temperature around 2.4~K. Although the
origin of this slight change of magnetization is unclear, it
should be noted that this temperature coincides with the onset of
a non trivial dynamical magnetic behaviour recorded in the ac
susceptibility measurements \cite{Minguet02}.

The analysis of the magnetic signal was made using the differences
between the 1.5 K and 10 K intensities of 52 reflections including
a set of ($h$,0,0) and (0,0,$l$) reflections with odd $h$ and $l$
indices. For these special reflections, the magnetic signal was
directly deduced from the integration of 10 mn/point
$\omega$-scans. Due to the weakness of the magnetic signal
compared to the nuclear one, for the mixed nuclear and magnetic
reflections, the magnetic signal was obtained from the difference
between 21 mn neutron counts at the peak maximum position
corrected from the side background, measured at 10 and 1.5 K. The
difference was then renormalized by the integrated intensity of a
3 mn/point $\omega$-scan.

To simplify the analysis, group theory can be used in order to
predict the possible magnetic arrangements compatible with the
crystal symmetries and as a consequence to reduce the number of
independent parameters in the refinement procedure. In the Landau
theory, the magnetic fluctuations in the paramagnetic state
reflect the crystal symmetries. At a second-order phase
transition, one of the modes is stabilized. Each mode corresponds
to one irreducible representation of the symmetry group $Gk$ that
leaves the propagation vector ${\bm k}$ invariant. The application
of the Group theory to the present crystal, using the tabulated
irreducible representations by Kovalev \cite{Kovalev}, shows that
the group representation can be reduced to a sum of 4 irreducible
representations of order 1 :
$\Gamma=3\tau^1\oplus3\tau^2\oplus3\tau^3\oplus3\tau^4$. The 4
possible magnetic arrangements deduced from this analysis are
summarized in Table \ref{tabtau}.

It is interesting to notice that the presence of a magnetic signal
on the ($h$,0,0) and (0,0,$l$) Bragg positions with odd $h$ and
$l$ allows to distinguish between these four magnetic structures.
$\tau^1$ is the only representation predicting a magnetic signal on
both ($h$,0,0) and (0,0,$l$) reflections. The $\tau^3$
representation yields no signal on both types of reflections. In
the case of $\tau^2$, a magnetic signal is expected on the ($h$,0,0)
reflections and not on the (0,0,$l$) ones, and the inverse is
predicted for the $\tau^4$ representation. For the present
measurements, the magnetic signal observed at 1.5 K on the
($h$,0,0) reflections with $h$=1,3,5 and on the (0,0,$l$)
reflections with $l$=3,5,7 therefore suggests a magnetic
arrangement associated with the $\tau^1$ representation.

\begin{table}
\begin{tabular}{|c|*{4}{c|}}  \hline
& $\tau^1$ & $\tau^2$ & $\tau^3$ & $\tau^4$  \\ \hline Mn &
1~2~3~4 & 1~2~3~4 & 1~2~3~4 & 1~2~3~4 \\ \hline $m_x$& +~-~+~- &
+~+~+~+ & +~+~-~- & +~-~-~+ \\ \hline $m_y$& +~-~-~+ & +~+~-~- &
+~+~+~+ & +~-~+~- \\ \hline $m_z$& +~+~-~- & +~-~-~+  & +~-~+~- &
+~+~+~+ \\ \hline
\end{tabular}
\caption{\label{tabtau} Magnetic components for the 4 Mn atoms :
Mn$_1$ (0.400, 0.189, 0.267), Mn$_2$ (0.100, 0.811, 0.767), Mn$_3$
(0.900, 0.311, 0.733) and Mn$_4$ (0.600, 0.689, 0.233) for each
irreducible representations $\tau^1$, $\tau^2$, $\tau^3$ and
$\tau^4$. }
\end{table}

\begin{figure}
\includegraphics[scale=0.45, bb=0 400 572 814]{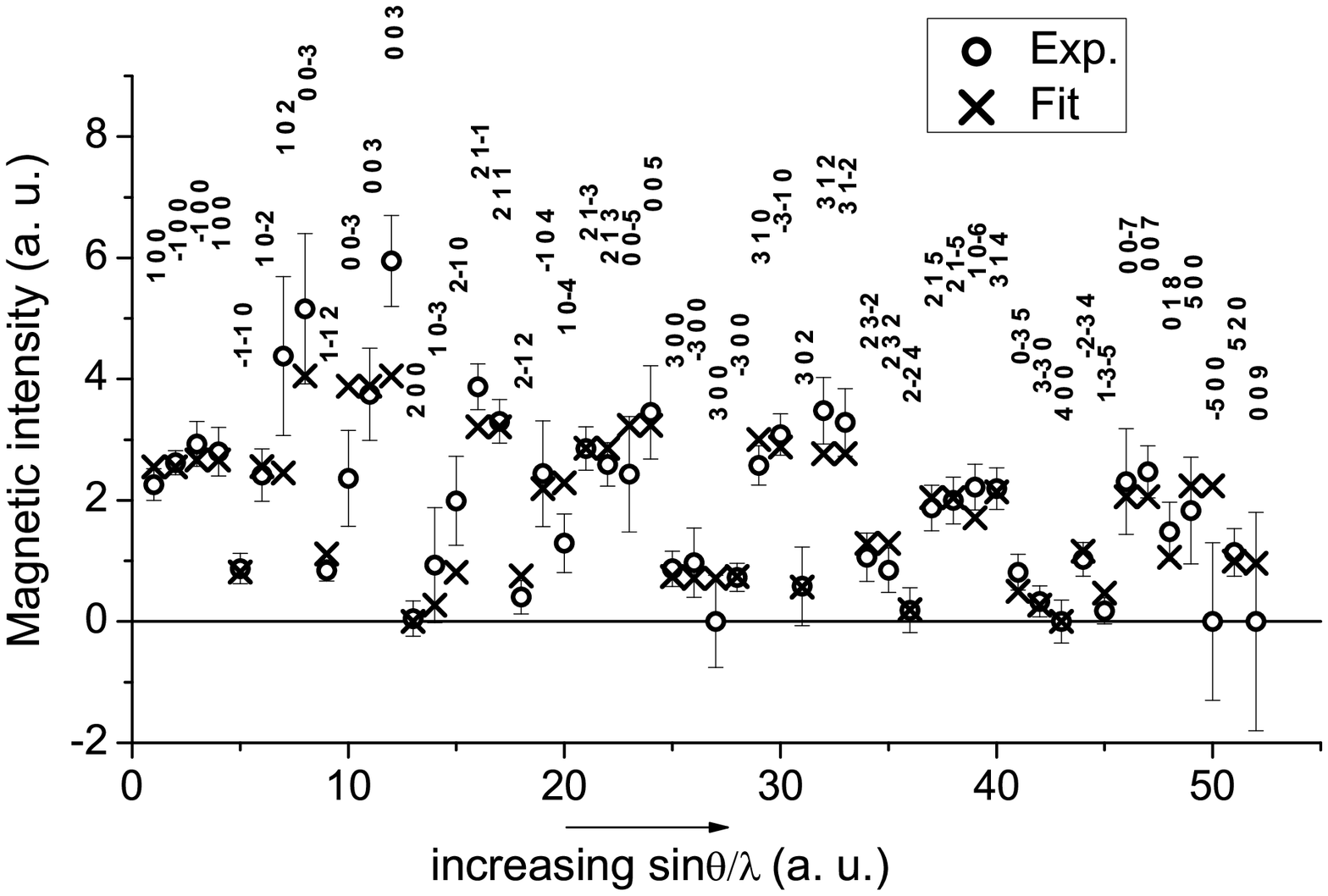}
\caption{Least square refinement of the magnetic structure in zero
field at 1.6 K: comparison of the measured and calculated
integrated intensities for the top labeled reflections displayed
with increasing $sin(\theta)/\lambda$.} \label{RefMag}
\end{figure}

The above representation analysis was carried out by considering
spins at the Mn positions only. However, a correct treatment of
the neutron data must in addition properly include the smaller
delocalized moment of the NITR free radicals which is strongly
antiferromagnetically coupled to the Mn one. If the distribution
of the magnetization (or form factor) around a Mn ion is well
known and tabulated, the major difficulty of evaluating the
radical contribution comes from the extended nature of the spin
distribution on this organic fragment. The magnetization
distribution in a quasi-isolated phenyl-substituted nytronil
nitroxide free radical was measured using polarized neutron
\cite{Zheludev94}. This work showed that, in such a radical,  most
of the spin density is found equally shared between the 4 atoms of
the two NO groups of the nitronyl nitroxide. An additional
negative contribution, due to the spin polarization effect, is
also observed on the central C atom. This picture has been used to
model the spin distribution on the nitronyl nitroxide in the
present case and to deduce a "pseudo" form factor that enters the
expression of the intensity of a magnetic reflection. In practice,
the magnetic structure factor was obtained at each scattering
vector $\bm{Q}$ by summing the contributions of the 4 Mn and the
4 NITR of the asymmetric unit, as in Eq. \ref{EqStrucF}. In this
equation, $f_{Mn}(Q)$ and $\bm{m}^j_{Mn}$ are respectively the
magnetic form factor and the magnetic moments for each Mn and
$p=0.2696 \times 10^{-12}$~cm. $\bm{m}_{NITR}^{j}$ is the magnetic
moment carried by the free radical whereas $(a_j+ib_j)$ is a
complex number evaluated at each measured ($h$,$k$,$l$) reflection
that represents both the "pseudo" form factor and the geometrical
term arising from the positions of the 5 atoms O, N, C, N and O
that are known to carry the spin in the radical.

\begin{equation}
\begin{split}
{\bm F}_M({\bm Q})=p\sum_{j=1}^{4}[&f_{Mn}(Q){\bm m}^j_{Mn}
\exp (i{\bm Q}.{\bm r}_j) \\
&+{\bm m}_{NITR}^{j}(a_j+ib_j)]
\label{EqStrucF}
\end{split}
\end{equation}

Note that a strong exchange interaction between the NITR and the
metallic centers can modify the spin distribution determined in
Ref. \onlinecite{Zheludev94,Schweizer}.  In our case this should
only slightly affect the quantitative results we have obtained.

\begin{figure}
\includegraphics[scale=0.5]{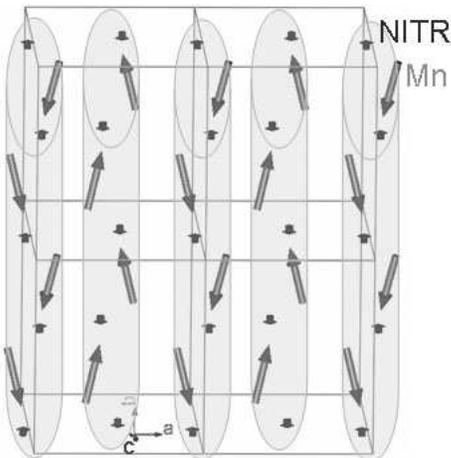}
\caption{Schematic representation of the magnetic structure at 1.5
K and zero magnetic field. The long and short arrows represent the
Mn and NITR moments respectively.} \label{MagH0}
\end{figure}

The final least-square refinement of the magnetic integrated
intensities with the $\tau^1$ representation for both Mn and NITR
radical was then performed. Note that the three other
representations yielded worse results ($\chi^2$=2.28, 2.05, 1.58
for $\tau^2$, $\tau^3$ and $\tau^4$ respectively) as expected from
the absence of magnetic signal on ($h$,0,0) and/or (0,0,$l$) Bragg
positions. The fit is also improved with respect to a calculation
with only an effective spin at the Mn position ($\chi^2$=2.03).
The best agreement (see Fig.~\ref{RefMag} and Table \ref{tabMag})
obtained with a reduced $\chi^2$ factor of 1.03 corresponds to a
canted antiferromagnetic structure (see Fig. \ref{MagH0}). Within
the chain, the main direction for the Mn$^{2+}$ moments is along
the $b$ axis, $m_y$=4.07$\pm$0.08 $\mu_{\rm B}$, with small
antiferromagnetic components along the $a$ and $c$ axis :
$m_x$=1$\pm$0.7 $\mu_{\rm B}$ and $m_z$=-0.7$\pm$0.5 $\mu_{\rm
B}$. The corresponding tilt angle with respect to the $b$ axis is
roughly equal to 16$^{\circ}$. There is a large uncertainty
concerning the exact orientation of the NITR moment but its main
component is along the chain axis, antiparallel to the Mn one.
Constraining the NITR moment to lie along the $b$ axis yields a
moment component of $m_y$=-0.7$\pm$0.1 $\mu_{\rm B}$. Note that
both Mn (4.25$\pm$0.3 $\mu_{\rm B}$) and NITR (0.7$\pm 0.1
\mu_{\rm B}$) moments have slightly reduced values. This could be
associated to some deterioration of the crystal or to a
delocalization of the NITR spin density on the Mn site. Finally,
the interchain coupling between the two chains of the unit cell is
found antiferromagnetic. The structure can then be described as
slightly canted ferromagnetic ($a$, $b$) sheets,
antiferromagnetically stacked along the $c$ direction with a
($a$/2, $b$/2) translation.

\begin{table}
\begin{tabular}{|c|ccc|ccc|}  \hline
& $x$ & $y$ & $z$ & M$_x$ & M$_y$ & M$_z$  \\ \hline\hline Mn$_1$
& 0.400 & 0.189 & 0.267 & 1.0 & 4.07 & -0.7 \\ \hline Mn$_2$&
0.100 & 0.811 & 0.767 & -1.0 & -4.07 & -0.7
\\ \hline Mn$_3$& 0.900 & 0.311 & 0.733 & 1.0 & -4.07 &
0.7 \\ \hline Mn$_4$& 0.600 & 0.689 & 0.233 & -1.0 & 4.07 & 0.7
\\ \hline\hline NITR$_1$& 0.480 & 0.440 & 0.230 & 0 & -0.7 & 0
\\ \hline NITR$_2$& 0.020 & 0.560 & 0.730 & 0 & 0.7 & 0
\\ \hline NITR$_3$& 0.980 & 0.060 & 0.770 & 0 & 0.7 & 0
\\ \hline NITR$_4$& 0.520 & 0.940 & 0.270 & 0 & -0.7 & 0
\\ \hline
\end{tabular}
\caption{\label{tabMag} Magnetic structure of Mn(hfac)$_2$({\it
R})-3MLNN in zero magnetic field at 1.5 K, refined assuming the
NITR moment to be parallel to the $b$ axis. The atomic coordinates
and components of the magnetic moments (in $\mu_{\rm B}$) are
given for the 4 Mn and NITR of the asymmetric unit. For the NITR,
the average magnetic moment is defined as in Eq. \ref{EqStrucF}
and the atomic coordinates are those of the central carbon atom of
the O-N-C-N-O fragment carrying most of the spin density. The
magnetic moment error bars are 0.7, 0.08 and 0.5 $\mu_{\rm B}$ for
the 3 components of the Mn and 0.1 $\mu_{\rm B}$ for the NITR $y$
component.}
\end{table}

\subsection{Evolution with a magnetic field}

The antiferromagnetic structure under zero field described in the
previous section is a surprising result, in apparent contradiction
with the magnetization and susceptibility measurements. This
motivated a subsequent neutron scattering experiment performed on
the D23 spectrometer during which the robustness of the magnetic
structure determined in zero field was tested under a magnetic
field applied parallel to the $b$ axis. A few peaks, in particular
the ($h$,0,0) and (0,0,$l$) ones with odd indices, characteristic
of the antiferromagnetic arrangement, were followed under a
magnetic field at 1.6~K.

\begin{figure}
\includegraphics[scale=0.4]{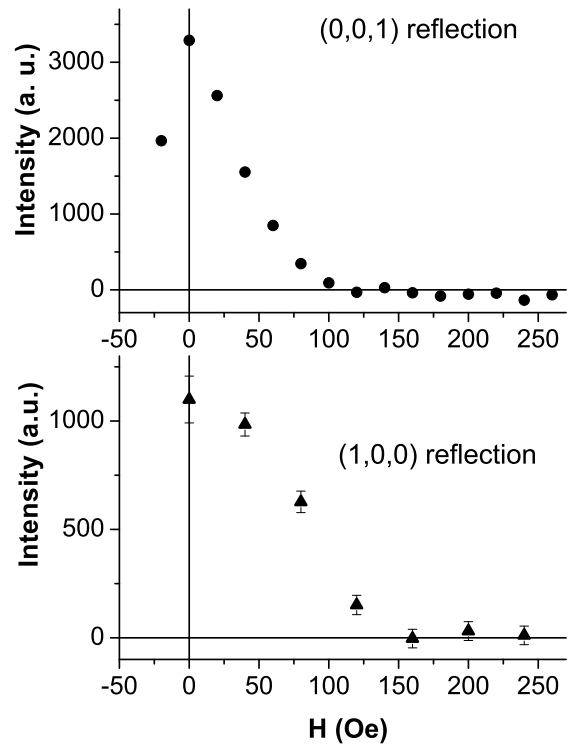}
\caption{Variation with magnetic field of the magnetic signal
associated to the (-1,0,0) and (1,0,0) reflections at 1.6 K :
neutron counts at the maximum peak position corrected from the
background. The error bars are within the symbol size for the
(0,0,1) reflection.} \label{MagH}
\end{figure}

The presence of the magnetic arrangement associated to the
$\tau^1$ representation was first confirmed by the rise of
magnetic signal observed on the (0,0,3), (0,0,1) and (1,0,0)
reflections between 5 and 1.6 K in zero field. At 1.6 K, the
magnetic intensities recorded on the (1,0,0) and (0,0,1)
reflections were then measured while varying the field and shown
to be wiped out very rapidly : A small field of the order of 100
to 150 Oe is sufficient to completely kill these magnetic
reflections (see Figs. \ref{MagH}). Note that a small hysteresis
was observed ($\sim$ 30 Oe shift), maybe due to the lack of
precision of the 6 T magnet in this range of applied fields.
Although the full evolution of the magnetic structure under a
magnetic field was not measured, the persistence of a magnetic
signal at a field $H_c$ sufficient to cancel the (0,0,$l$) and
($h$,0,0) reflections was checked. This was achieved by measuring
the difference between the integrated intensities at 5 K in zero
field and at 1.6 K in $H_c$ on both (-2,1,0) and (1,0,-4)
reflections. The magnetic signal is significantly reduced on
(1,0,-4) and increased on (-2,1,0) when applying the field. This
is in agreement with the expected field-induced transition towards
a ferromagnetic interchain arrangement of the moments ($\tau^3$
representation), possibly slightly canted although the accuracy of
the measurement is not sufficient to quantify this point.

\section{Discussion}

The antiferromagnetic LRO determined by neutron diffraction below
$T_c$ is therefore different from the one expected through the
observation at higher temperatures of 3D ferromagnetic
correlations. It is furthermore very easily destabilized under a
weak magnetic field. To interpret this behavior, we will consider
in the following the magnetic couplings that are effective in this
temperature range, and in particular their interchain
contribution, as a necessary ingredient to drive the system to a
3D LRO.

A good candidate for this interchain mechanism, often invoked in
chain compounds \cite{CaneschiMn,Walker,Wynn,Ostrovsky01}, is the
weak but long-ranged dipolar coupling. In addition to its role in
the onset of 3D LRO, the dipolar interaction is also argued to be
responsible for the magnetic anisotropy observed in some Mn-based
chain compounds \cite{CaneschiMn,Walker}. The importance of
dipolar anisotropy is a consequence of the weakness of the
single-ion anisotropy, describable as a crystal zero field
splitting of the ion ground state. For Mn$^{2+}$, there is no
zero-field splitting in 1$^{st}$ order of perturbation due to the
zero orbital angular momentum of its $^6$S$_{5/2}$ ground state.
Non-zero values can only arise from higher order processes
allowing the interaction of the ions with the crystal field, the
most important mechanism being the admixtures through spin-orbit
coupling of $L\not = 0$ excited states within the ground state
\cite{Edgar80}. In the present Mn-NITR chain compound, we observe
a weak axial anisotropy of the magnetic moments. They are mainly
oriented along the chain axis, with a small canting. Such a
magnetic orientation appears to be rare in spin chains where the
moments usually lie perpendicular to the chain axis. This
perpendicular orientation is expected for systems consisted of an
antiferromagnetic stacking of spins along the chain axis in which
the dipolar interaction is the main source of anisotropy. The
origin of the anisotropy in the present case where the magnetic
entities are Mn$^{2+}$ and organic radicals with isotropic spins
1/2 is therefore questionable.

In order to check the role of dipolar coupling  in the
orientation of the spins and in the onset of a finite temperature LRO, we have
used a mean field (MF) approach. The method is only briefly
described in this paper but more details about the formalism may
be found for instance in Ref.~\onlinecite{Enjalran2004}. We have
used a general anisotropic Hamiltonian which takes into account
the intrachain isotropic exchange couplings ($J=-510$ K) between nearest
neighbor Mn ($\mu_{\rm Mn} = 5/2$) and NITR radicals ($\mu_{\rm NITR} =
 1/2$), the long range anisotropic dipolar
coupling between all species of magnetic atoms, as well as a local
anisotropy on the manganese sites. The detail of the calculations
are reported in the Appendix.

The role of dipolar interactions was first tested by neglecting
the single-ion anisotropy term, $D=0$. The resulting
antiferromagnetic LRO was found with the propagation vector ${\bm
k}$=0, as observed experimentally. However, contrary to the
anisotropy inferred from neutron scattering and magnetometry
measurements, the spin orientation was found along the $a$ axis,
i.e. perpendicular to the chain direction. Consequently, the
influence of the single-ion anisotropy was evaluated by
introducing a local uniaxial anisotropy $D > 0$ whose value was
increased progressively.
As results, the antiferromagnetic structure with ${\bm k}$=0
remains for all $D$ values, but the spins direction switches to
the $b$ axis for $D/k_{\rm B}$ values greater than $\approx 160$~
mK. This is larger but of the same order of magnitude than the
value, $D/k_{\rm B}\approx$ 40 mK, estimated from the anisotropy
constant $K$ obtained via the low temperature magnetization
measurements \cite{footnote4}. Note that due to some
approximations made in the calculations (the interchain exchange
coupling and the small canting with respect to the $b$ axis were
not taken into account for instance) the results should be
considered as qualitative only.

These results indicate first that the dipolar interaction cannot
explain the anisotropy observed in the chain, contrary to the case
of other Mn-NITR chains \cite{CaneschiMn}. Other mechanisms must
then be invoked such as antisymmetric exchange or Mn$^{2+}$
single-ion anisotropy. In support of this last explanation, let us
note that, although often negligible, large zero field splittings
have been reported in the Cs$_2$MnCl$_4$.2H$_2$0 antiferromagnet
($|D/k_{\rm B}|\sim$140 mK) \cite{smith69} for instance and in
several Mn(II) organic compounds ($|D/k_{\rm B}|$ up to $\sim$1 K)
\cite{Lynch93}. The single-ion anisotropy has also been proposed
to be the main source of anisotropy in other Mn(II) chain
compounds \cite{Luneau1, smith68}.  In fact, one of these
compounds undergoes a transition towards an antiferromagnetic LRO
and is one of the rare example with the direction of the Mn
moments along the chain axis \cite{smith68}.

The second important result of these MF calculations is that the
interchain dipolar interaction produces a 3D antiferromagnetic LRO
with propagation vector ${\bm k}$=0, which is robust with respect
to the spins orientation. This interaction is therefore most
probably responsible for the antiferromagnetic LRO observed in
Mn(hfac)$_2$({\it R})-3MLNN below $T_c$.

Nevertheless, it seems difficult to reconcile the amplitude of the
dipolar interaction between two spins in neighboring chains,
$d_{nn}$, and the critical temperature: $d_{nn}=\displaystyle
(\mu_0 / 4 \pi) \, (g \mu_{\rm B})^2 \, / \, r_{nn}^3 \approx 1.7$
mK, for a nearest neighbor chain distance of $r_{nn}=11.33$~\AA,
which is three orders of magnitude smaller than $T_c$=3 K.

As shown below, this apparent contradiction can be resolved as
soon as one considers the intrachain correlations developing at
low temperatures, which build super spins made of a temperature
dependent number of strongly correlated individual spins.

The critical temperature can be calculated in a mean field
approach, considering only intrachain exchange and interchain
dipolar interaction. To describe the low temperature behavior of
quantum-classical ferrimagnetic chains in the Heisenberg model,
let us consider a single chain as independent quasi-rigid blocks,
or super spins, of length $\xi$ and total moment $\tilde S=\xi
S_{tot}$ with $S_{tot}=S-s=2$ ($S$=5/2 and $s$=1/2). The
correlation length is then $\xi=|\tilde J|/2 k_B T$ with $\tilde
J=JSs$ ($J/k_{\rm B}$=-510~K the intrachain correlation)
\cite{Curely95}.

As we know from experiments that these super spins tend to align
along the $b$ axis of the crystal, we end up with an effective
model of pseudo Ising super spins $\tilde S$ distributed on a
distorted triangular lattice and interacting via the dipolar
interaction. Using $d_{nn}/k_B$=1.7 mK in the previously introduced formalism, the mean-field critical temperature is given by

\begin{equation}
T_c^{\rm MF}= \frac{c}{k_B}~d_{nn} \tilde S^2 /n=\frac{c}{k_B}~d_{nn} (\xi(T_c)S_{tot})^2 /n
\label{xi2}
\end{equation}
where $n=3$ is the spin dimension and $c\approx 1.9$ is a
dimensionless constant dependent on the two dimensional super spin
lattice geometry. This finally leads to :
\begin{equation}
T_c^{\rm MF}=\frac{1}{k_B} \left(\frac{c}{4n}~d_{nn}~S_{tot}^2~|\tilde J|^2 \right)^{1/3}
\label{eqTc}
\end{equation}
from which we obtain $T_c^{\rm MF} \simeq 7.6$ K.
Keeping in mind that the mean field approach always tends to
overestimate the transition temperature, this result seems in
rather good agreement with the measured $T_c=3$~K.
Perhaps even more important is that this calculation establishes that the
LRO is driven both by the intrachain exchange interaction
which operates first and builds mesoscopic super spins (at
$T_c^{\rm MF}$, around 80 spins are strongly correlated) and by
the dipolar interchain interaction which settles the 3D ordering
temperature of these super spins.

As argued, the dipolar interaction should be responsible for the
3D antiferromagnetic stacking of the chains.
Consequently, it cannot explain the 3D ferromagnetic correlations
observed above $T_c$. These correlations must then be induced by
the interchain exchange interaction in spite of the large
interchain distances. Indeed, although short-range in nature, the
exchange interaction in molecular compounds can be mediated by the
spin delocalization and polarization processes on the whole NITR
molecule (not only on the two O-N bridges of the radical) and
between several molecules \cite{Schweizer}. In the present
structure, some C-H$\dotsb$O interchain paths of lentgh 2.7~\AA~ can be identified,
which fall within the range of weak hydrogen bonds
\cite{Alonso04}. Hydrogen bonds have been shown to be possible
exchange paths between molecules \cite{Romero00}. Such exchange
mechanism should lead to a small value of the interaction energy,
which is consistent with the value of the interchain exchange
obtained from the mean field approach ($J_{inter}/k_{\rm B}
\approx~$4~mK, see Section II) and of the same order as the
dipolar energy. However, the sign of this interaction cannot be
simply deduced from structural consideration due to the complexity
of the exchange paths. Note that previous magnetization
measurements under pressure on Mn(hfac)$_2$({\it
R})-3MLNN also concluded the existence of ferromagnetic
interchain exchange \cite{Laukhin04}. Polarized neutron
experiments on a single-crystal could give more information on these interactions.

The unusual transition revealed by these experiments
(ferromagnetic-like susceptibility and antiferromagnetic LRO)
could therefore result from the competition between the
ferromagnetic interchain exchange and the antiferromagnetic
dipolar interaction, and from their different sensitivity to the
strong 1D magnetic behavior. Decreasing the temperature leads to
the formation within the chains of correlated blocks of spins
whose length $\xi$ increases like the inverse of the temperature.
The long-range nature of the dipolar interaction makes it more
sensitive to the intrachain correlations than the exchange
interaction \cite{Ostrovsky01}. This is reflected by the
transition temperature obtained in a mean field approach: $T_c$
increases like $\xi$ for the exchange interaction
\cite{Villain77}, whereas it increases like $\xi^2$ for the
dipolar one (Eq. \ref{xi2}). Therefore, below a certain
temperature, when $\xi$ reaches a threshold value, the dominant
exchange interaction leading to ferromagnetic interchain
correlations can be overwhelmed by the dipolar one. This in turn
should favor antiferromagnetic interchain arrangements.

This interpretation of the magnetic properties of the
Mn(hfac)$_2$({\it R})-3MLNN compound in terms of competition
between weak exchange and dipolar interactions is supported by the
magnetic behavior under weak magnetic fields. In the M(H) curves
(see Section II), a small curvature appears at fields of a few Oe
and the intensity of magnetic Bragg peaks associated with the
antiferromagnetic order starts to decrease under the same field of
a few Oe (see Section III). These observations are the signature
of a fast deterioration of the zero-field antiferromagnetic ground
state in favor of a ferromagnetic LRO under a small field. They
are consistent with the estimated strength of the
antiferromagnetic dipolar interchain interaction which leads to an
equivalent field of a few Oe. This transition is however not
characterized by an abrupt metamagnetic transition. This could be
explained by a distribution of the intrachain correlation lengths
$\xi$, which is especially relevant in such compounds due to the
finite size of the chains \cite{Bogani04}. Because of the
particular sensitivity of the dipolar interaction to the
correlation length, a distribution of the $\xi$ values would
result in a distribution of the reversal fields instead of a
unique field characteristic of a well-defined metamagnetic
transition.

It is interesting to note a problem reported in some previously
studied Mn-NITR chains \cite{Caneschi94} were  ferromagnetic LRO
was inferred from magnetization measurements, in contradiction to
dipolar coupling calculations on the same compounds that predicted
an antiferromagnetic structure. Furthermore, the calculations
showed that ferromagnetic LRO could only be stabilized in magnetic
fields higher than 20 Oe. In the present study, we measured
similar ferromagnetic signatures by magnetometry and in addition,
established the antiferromagnetic nature of the LRO by neutron
diffraction. The above analysis provides a scenario for the
coexistence of these two types of fluctuations and suggests that
the temperature ranges where each one dominates is driven by the
intrachain correlation length.

\section{Conclusion}
Magnetization measurements on the Mn-free radical chiral chain
Mn(hfac)$_2$({\it R})-3MLNN compound have shown ferromagnetic
correlations and a magnetic phase transition at $T_c=$3 K, but
with a puzzling behavior in the ordered regime. Neutron
measurements allowed to solve this intringuing problem.

In the present paper, the resolution of the magnetic structure of
this compound from single-crystal neutron diffraction measurements
was obtained using specific methods of analysis (rigid blocks
refinement, symmetry analysis, "molecular" magnetic form factors)
in order to overcome (i) the weakness of the signal and the fast
deterioration of this small and brittle molecular sample and (ii)
the presence of delocalized spins on the organic free radicals.
Under moderate fields (H$>$150 Oe), the existence of a
ferromagnetic LRO, evidenced by macroscopic magnetostatic studies,
is confirmed. However in zero magnetic field, the actual magnetic
LRO stabilized below $T_c$ is found to be an
antiferromagnetic arrangement of canted ferromagnetic chains in
contrast with the ferromagnetic 3D fluctuations observed above
$T_c$.

This result demonstrates that two types of weak interchain interactions,
of opposite signs, are in competition in this compound. As shown
by mean-field calculations, the dipolar one, enhanced by the
intrachain spin correlations, dominates below T$_c$ and leads to
an interchain antiferromagnetic structure that rapidly vanishes
under weak magnetic field. The weak exchange interaction is mainly influent
above T$_c$ producing strong ferromagnetic 3D correlations, a
ferromagnetic LRO being only field-induced below T$_c$.

These results underline the need, in the field of molecular
magnetism where weakly competing processes may be active, of using
both macroscopic magnetization measurements {\it and} microscopic
probes such as neutron diffraction.

\acknowledgments P. Wolfers is acknowledged for his help in
implementing new developments of his MXD program for handling
rigid block refinement. We thank A. Sulpice for his high
temperature magnetization measurements and B. Grenier for her help
during the neutron diffraction experiments. This work has been
supported in part by the European Commission under the Network of
Excellence MAGMANet (contract 515767-2), by the Ministerio de
Educaci\'on y Ciencia (Spain), under the project MAT2003-04699 and
the project ejeC-Consolider CTQ2006-06333/BQU, and by Generalitat
de Catalunya (2005SGR00362). Support from the R\'egion Rh\^one-Alpes  through the "Programme de Recherche Th\'ematiques Prioritaires" is also  gratefully acknowledged.

\appendix
\section{Mean field calculations}
\label{appendix}

For the mean field calculations\cite{Enjalran2004} of the magnetic
configuration selected at low temperature in presence of
intrachain exchange interaction, dipolar interaction and
single-ion anisotropy, we introduce the general anisotropic
Hamiltonian, $\mathcal{H}$:
\begin{eqnarray}
\mathcal{H} &=& - \frac{1}{2}\sum_{i,j} \sum_{m,n} \sum_{\alpha,
\beta}  {\mathcal{J} \left( R_{ij}^{mn} \right)}^{\alpha \beta}
 {\mu}_{i}^{m,\alpha}
{\mu}_{j}^{n,\beta} \nonumber \\ &&- D \sum_{(i,m)\equiv{\rm Mn}}
{{\mu}_{i}^{m,z}}^2 , \label{ham}
\end{eqnarray}
where
\begin{eqnarray}
\label{general-J} & &{\mathcal{J} \left( R_{ij}^{mn}
\right)}^{\alpha \beta}  = J  \delta_{R_{ij}^{mn},r_{nn}} \;
({\hat n}^{\alpha} \cdot {\hat n}^{\beta})  \\ & &- d_{nn} \,
{\mu}_i^n \, {\mu}_j^m \, \left ( \frac{({\hat n}^{\alpha}\cdot
{\hat n}^{\beta})}{|R_{ij}^{mn}|^3} - \frac{3 ({\hat
n}^{\alpha}\cdot R_{ij}^{mn}) ({\hat n}^{\beta}\cdot
R_{ij}^{mn})}{|R_{ij}^{mn}|^5} \right )\; . \nonumber
\end{eqnarray}
In the notation of this general model, the moment vectors are
represented by ${\bm \mu}_i^m = {\mu}_i^m ( {\hat
n}^{(x)}{S}_i^{m,x}+{\hat n}^{(y)}S_i^{m,y}+ {\hat
n}^{(z)}S_i^{m,z})$, where the unit vectors ${\hat n}^{\alpha}$
are the global Cartesian unit vectors, and $S_i^{m,\alpha}$ is the
$\alpha$-th component of a unitary spin \cite{footnote2}.
${\mu}_i^m$ is the moment of the atom residing at site $(i,m)$.
$J$ is the nearest neighbor exchange interaction ($J<0$ for an
antiferromagnet), $r_{nn}$ the nearest neighbor distance, $D$ the
anisotropy, taken as uniaxial for simplicity and $d_{nn}$ the
nearest neighbor dipolar coupling, $$ d_{nn} = \frac{\mu_0}{4 \pi}
\frac{(g \mu_{\rm B})^2}{r_{nn}^3} $$
Two indices are needed to localize a site as we describe the
lattice as a Bravais lattice with eight sites per unit cell.
It must be noted that the sum in Eq.~\ref{ham} does not include
terms with ${\bm R}_{ij}^{mn}=0$ and that in Eq.~\ref{general-J},
positions ${\bm R}_{ij}^{mn}$ are taken in units of $r_{nn}$.

Before writing the mean-field free energy, the Hamiltonian is
first rewritten in reciprocal space using the following
transformations,
\begin{eqnarray}
\label{eq-FTm} S_{i}^{m,\alpha} = \frac{1}{\sqrt{N}} \sum_{\bm q}
S_{\bm q}^{m,\alpha} \, e^{-\imath {\bm q} \cdot {\bm R}_{i}^{m}},
\\ {\mathcal{J} \left( R_{ij}^{mn} \right)}^{\alpha \beta} =
\frac{1}{N} \sum_{\bm q} {\mathcal J}_{mn}^{\alpha \beta}({\bm q})
\, e^{\imath {\bm q} \cdot {\bm R}_{ij}^{mn}}, \label{eq-FTJ}
\end{eqnarray}
where $N$ is the number of Bravais lattice points.
Because we deal with a non Bravais lattice, which is moreover non
centro symmetric, the resulting interaction matrix ${\mathcal
J}({\bm q})$ is a $24 \times 24$ non diagonal hermitian matrix.
Hence, to completely diagonalize ${\mathcal J}({\bm q})$ one must
transform the ${\bm q}$-dependent variables, ${\bm S}_{\bm q}^m$,
to normal mode variables. In component form, the normal mode
transformation is given by
\begin{equation}
\label{eq-nmodes} S_{\bm q}^{n,\alpha} =
\sum_{p=1}^{8}\sum_{\gamma=1}^{3} U_{n,p}^{\alpha,\gamma}({\bm q})
\phi_{\bm q}^{p,\gamma},
\end{equation}
where the indices ($p,\gamma$) label the normal modes ($24$ for
Heisenberg spins), and $\{\phi_{\bm q}^{p, \gamma}\}$ are the
amplitudes of these normal modes. $U({\bm q})$ is the unitary
matrix that diagonalizes ${\mathcal J}({\bm q})$ with eigenvalues
$\lambda({\bm q})$.
Finally, the mean-field free energy to quadratic order in the
normal modes reads, up to an irrelevant constant,
\begin{equation}
{\mathcal F}(T) = \frac{1}{2} \sum_{{\bm q}, p, \gamma} (nT -
\lambda_{p}^{\gamma}({\bm q})) |\phi_{\bm q}^{p, \gamma}|^2 ,
\label{eq-fmf}
\end{equation}
where ${\mathcal F}(T)$ is the mean-field free energy per unit
cell, $T$ is the temperature in units of $k_{\rm B}$, and $n=3$
for Heisenberg spins.

Therefore, the mean-field critical temperature is given by $$
T_c^{\rm MF} = \frac{1}{n} {\max}_{p, \gamma, {\bm q}} \left(
\lambda_{p}^{\gamma}({\bm q}) \right) $$ where the corresponding
wave vector ${\bm q}_{\rm ord}$ defines the magnetic propagation
vector ${\bm k}$.
If the extremal eigenvalue is non degenerate (which is always the
case in this study), the inner structure of the magnetic unit cell
is given by the corresponding eigenvector ${\bm U}_p^{\gamma}({\bm
q}_{\rm ord})$.

The main technical difficulty arises because the interaction
matrix is obtained from a sum which is conditionally convergent
due to the long range interaction of the dipolar coupling.
The usual method to overcome this difficulty is to use the Ewald's
method\cite{Ewald1921}.
It appeared that this is not necessary in the present case.
Whatever the truncation radius we took, as soon as it is large
enough \cite{footnote3}, we found that the ordering wave vector is
always stable and that there were only slight modifications on the
corresponding eigenvector.
We therefore restricted ourselves to a finite sum in direct space.


\begin{thebibliography}
\bibitem{footnote1} This value obtained using the Seiden model \cite{Seiden83} has
been refined compared to ref. \onlinecite{Minguet02} thanks to
more precise measurements.
\bibitem{footnote4} The $D$ value is calculated for single-ion anisotropy on the Mn site only, whereas the
$D$ estimated from the measurements is derived for a model of
$S_{tot}$=2 magnetic entities. These values are therefore not
directly comparable but should be in qualitative agreement.
\bibitem{footnote2} Note that the dipolar term in $\mathcal{H}$ was computed with either a
localized spin 1/2 on the NITR average position or with fractions
of spin on both N-O fragments of the NITR in order to take into
account the delocalizion of the spin. Similar results were found
in both calculations.
\bibitem{footnote3} Beyond a truncation radius of 300 times the nearest neighbor
distance, even the eigenvector modifications are not noticeable,
e.g, lower than 10$^{-2}$ for a normalized eigenvector.
\end{thebibliography}
\end{document}